\documentclass[letter]{aa} % for the letters 

\usepackage{graphicx}
\usepackage{txfonts}
\usepackage{natbib,twoopt}
\usepackage[breaklinks=true]{hyperref} %% to avoid \citeads line fills
\bibpunct{(}{)}{;}{a}{}{,}             %% natbib format for A&A and ApJ
\newcommand{\lya}{Ly$\alpha$}

\def\lesssim{\mathrel{\hbox{\rlap{\hbox{%
 \lower4pt\hbox{$\sim$}}}\hbox{$<$}}}}
\def\gtrsim{\mathrel{\hbox{\rlap{\hbox{%
 \lower4pt\hbox{$\sim$}}}\hbox{$>$}}}}

\begin{document} 

   \title{Energy-limited escape revised{}}

   \subtitle{The transition from strong planetary winds
             to stable thermospheres\thanks{
          Appendices are available in electronic form.}}

   \author{M. Salz\inst{1},
           P. C. Schneider\inst{2}$^,$\inst{1},
           S. Czesla\inst{1},
           J. H. M. M. Schmitt\inst{1}
          }

   \institute{Hamburger Sternwarte, Universit\"at Hamburg,
               Gojenbergsweg 112, 21029 Hamburg, Germany\\
              \email{msalz@hs.uni-hamburg.de}
         \and
              European Space Research and Technology Centre (ESA/ESTEC),
              Keplerlaan 1, 2201 AZ  Noordwijk, The Netherlands
             }

   \date{Received 23 July 2015 / Accepted 10 November 2015}
   
   \abstract{
     Gas planets in close proximity to their host stars experience
     photoevaporative mass loss.
     The energy-limited escape concept is generally used to derive
     estimates for the planetary mass-loss rates. 
     Our photoionization hydrodynamics simulations of the thermospheres of
     hot gas planets show that the energy-limited escape concept is
     valid only for planets with a gravitational potential lower than
     $\log_\mathrm{10}\left( -\Phi_{\mathrm{G}}\right) < 13.11$~erg\,g$^{-1}$
     because in these planets the radiative energy input is efficiently
     used to drive the planetary wind. 
     Massive and compact planets with
     $\log_\mathrm{10}\left( -\Phi_{\mathrm{G}}\right) \gtrsim 13.6$~erg\,g$^{-1}$
     exhibit more tightly bound atmospheres in which the complete radiative
     energy input is re-emitted through hydrogen \lya{} and free-free emission.
     These planets therefore host hydrodynamically stable thermospheres.
     Between these two extremes the strength of the planetary winds rapidly declines as a result of a decreasing heating efficiency.
     Small planets undergo enhanced evaporation because they host expanded atmospheres that expose a larger surface to the stellar irradiation.
     We present scaling laws for the heating efficiency and the expansion radius that depend on the gravitational potential and irradiation level of the planet.
     The resulting revised energy-limited escape concept can be used to derive estimates for the mass-loss rates of super-Earth-sized planets as well as massive hot Jupiters with hydrogen-dominated atmospheres.
   }


   \maketitle

%_______________________________________________________________________________
%_______________________________________________________________________________
\section{Introduction}\label{SectIntro}
%_______________________________________________________________________________
%_______________________________________________________________________________

More than 30 years ago, \citet{Watson1981} demonstrated that planetary atmospheres can become hydrodynamically unstable when exposed to strong high-energy irradiation. Therefore, the discovery of hot Jupiters \citep{Mayor1995} soon raised questions about the stability of their atmospheres \citep[e.g.,][]{Lammer2003}. In particular, the absorption of high-energy radiation causes ionizations with subsequent thermalization of the kinetic energy of photoelectrons. The resulting heating initiates a process of continuous thermospheric expansion, thus, launching a planetary wind powered by the stellar high-energy emission \citep[e.g.,][]{Yelle2004, Tian2005, Garcia2007, Penz2008-2, Murray2009, Koskinen2013a, Shaikhislamov2014}.

Today, the impact of atmospheric mass loss on the evolution of hot Jupiters is believed to be small because the fractional mass loss remains moderate throughout a planetary lifetime \citep[e.g.,][]{Ehrenreich2012}. However, photoevaporative mass loss can play a more decisive role for the evolution of smaller gas or terrestrial planets. \citet{Salz2015c} demonstrated that planets like 55\,Cnc\,e may lose all volatiles over a planetary lifetime, eventually leaving a rocky super-Earth-type core behind. In fact, photoevaporative mass loss may be among the crucial factors determining the distribution of small close-in planets \citep{Lecavelier2007, Carter2012}.

Expanded planetary thermospheres that probably undergo hydrodynamic escape have been reported for the five systems HD\,209458\,b, HD\,189733\,b, WASP-12\,b, 55\,Cancri\,b, and GJ\,436\,b, mostly based on \lya{} transit spectroscopy \citep{Vidal2003, Lecavelier2010, Fossati2010, Ehrenreich2012, Kulow2014}. In principle, stellar wind interactions or strong planetary magnetic fields can inhibit the formation of freely escaping planetary winds \citep{Murray2009, Trammell2014}, nevertheless, the mass-loss rates of hot gas planets are often approximated by applying the idealized energy-limited escape model.

%_______________________________________________________________________________
%_______________________________________________________________________________
\vspace{-4mm}
\section{Energy-limited escape}\label{SectErgLimEsc}
%_______________________________________________________________________________
%_______________________________________________________________________________

The concept of energy-limited escape is based on an energy budget consideration that treats the planetary thermosphere as a closed system \citep[e.g.,][]{Watson1981, Erkaev2007}. A mass-loss rate is derived by setting the radiative energy input equal to the total specific energy gain of the evaporated atmospheric material.

%_______________________________________________________________________________
\vspace{-3mm}
\subsection{Energy-limited escape equation}\label{SectErgLimEq}
%_______________________________________________________________________________

Considering only the gain of potential energy of the evaporated material and neglecting kinetic and thermal terms, the following formula can be derived for the planetary mass-loss rate \citep[e.g.,][]{Erkaev2007, Sanz2010}:
\vspace{-2mm}
\begin{equation}\label{EqMloss_erglim}
  \dot{M}_{\mathrm{el}} = \frac{3\,\beta^2\,\eta F_{\mathrm{XUV}}}{4\,KG\,\rho_{\mathrm{pl}}} \, .
\end{equation}

\vspace{-3mm}\noindent Here, $\rho_{\mathrm{pl}}$ is the mean planetary density and $G$ the gravitational constant. The radiative energy input is given by the stellar high-energy flux $F_{\mathrm{XUV}}$ (X-ray plus extreme UV irradiation) multiplied by the area of the planetary disk. $\beta = R_{\mathrm{XUV}}/R_{\mathrm{pl}}$ is a correction factor for the size of the planetary disk that absorbs XUV radiation\footnote{
Equation~\ref{EqMloss_erglim} is sometimes given with a factor of $\beta^3$ \citep[e.g.,][]{Baraffe2004, Sanz2010}, but we favor $\beta^2$ \citep{Watson1981, Lammer2003, Erkaev2007}.
The equation contains $R^3_{\mathrm{pl}}$ in $\rho_{\mathrm{pl}}$, but one of the radii comes in via the gravitational potential. Correcting the potential with $\beta$ implies that the material escapes from $R_{\mathrm{XUV}}$, but the energy to lift material from $R_{\mathrm{pl}}$ to $R_{\mathrm{XUV}}$ must also be provided.
};
$R_{\mathrm{XUV}}$ is the mean XUV absorption radius.
The heating efficiency $\eta$ specifies the fraction of the radiative energy input that is available for atmospheric heating. In particular, a certain fraction of the energy input is used for ionizations, and in addition, recombinations and collisional excitations cause radiative cooling that also reduces the available energy. Atmospheric material only has to escape to the Roche-lobe height, $R_{\mathrm{Rl}}$, which is accounted for by multiplying with the fractional gravitational potential energy difference, $K$, between the planetary surface and the Roche-lobe height \citep{Erkaev2007}:
\vspace{-1mm}
\begin{equation}\label{EqK_factor}
  K(\xi) = 1 - \frac{3}{2\xi} + \frac{1}{2\xi^3} \quad\mathrm{with}\quad
  \xi = \left(\frac{M_{\mathrm{pl}}}{3M_{\mathrm{st}}}\right)^{1/3} 
  \frac{a}{R_{\mathrm{pl}}}  .
\end{equation}
Here, $M_{\mathrm{pl}}$ and $M_{\mathrm{st}}$ are the planetary and stellar masses, $a$ is the semimajor axis, and $R_{\mathrm{pl}}$ is the planetary radius. This approximate expression is valid for $a\gg R_{\mathrm{Rl}}>R_{\mathrm{pl}}$ and $M_{\mathrm{st}}\gg M_{\mathrm{pl}}$.

%_______________________________________________________________________________
\vspace{-3mm}
\subsection{Uncertainties in the energy-limited escape concept}\label{SectErgLimWeak}
%_______________________________________________________________________________

Energy-limited escape is either used for upper limits of planetary mass-loss rates, or for order-of-magnitude estimates by choosing an appropriate heating efficiency \citep[e.g.,][]{Lecavelier2007, Ehrenreich2011,2015Salza}. 
While numerical studies have shown that energy-limited escape rates can be reached in the atmospheres of hot Jupiters \citep[e.g.,][]{Garcia2007, Murray2009},
there are three questions that give rise to substantial uncertainty.
The first is the use of an a priori unknown heating efficiency for the absorption of XUV radiation. In the literature, values from 0.1 to 1.0 have been adopted for this essential factor \citep{Shematovich2014}, but we show that heating efficiencies vary by several orders of magnitude in individual atmospheres.
The second question is the unknown size of the planetary atmosphere that absorbs the XUV radiation. We demonstrate that the atmospheric expansion increases the mass-loss rates of smaller planets \citep[cf.,][]{Erkaev2013}.
The third question is the general assumption of hydrodynamic escape, which is only valid if a planetary thermosphere is collisional up to the Roche-lobe height or the sonic point. Radiative heating cannot be converted into a bulk acceleration above the exobase, where the planetary atmosphere becomes collisionless; our definition of the exobase is given by \citet{Salz2015c}. In this case, a combined hydrodynamic and kinetic description has to be applied \citep[e.g.,][]{Yelle2004}, which results in reduced mass-loss rates ranging between the Jeans escape rate and the hydrodynamic escape rate \citep[cf. Fig.~13 of][]{Tian2005}. Therefore, the energy-limited concept cannot readily be applied to collisionless exospheric escape.

%_______________________________________________________________________________
%_______________________________________________________________________________
\vspace{-2mm}
\section{Simulations: The transition from hydrodynamic winds to stable thermospheres}\label{SectSims}
%_______________________________________________________________________________
%_______________________________________________________________________________

 \citet{Salz2015c} presented simulations of the thermospheres of 18 hot gas planets in the solar neighborhood using 1D spherically symmetric simulations of the substellar point with an atmospheric composition of hydrogen and helium. The simulations were carried out with the PLUTO-CLOUDY interface \citep[TPCI,][]{Salz2015b}. This is an interface between the magnetohydronynamics (MHD) code PLUTO \citep{Mignone2012} and the photoionization solver CLOUDY \citep{Ferland2013}. Since CLOUDY solves the relevant microphysical processes from first principles, we can use our simulations to derive the heating efficiencies in the individual atmospheres. TPCI can be applied to rapidly escaping atmospheres, where the radiative energy input is almost completely used to drive the wind, but also to hydrostatic atmospheres, where the absorbed energy is re-emitted. 

\begin{figure}[tb]
  \centering
  \includegraphics[width=\hsize]{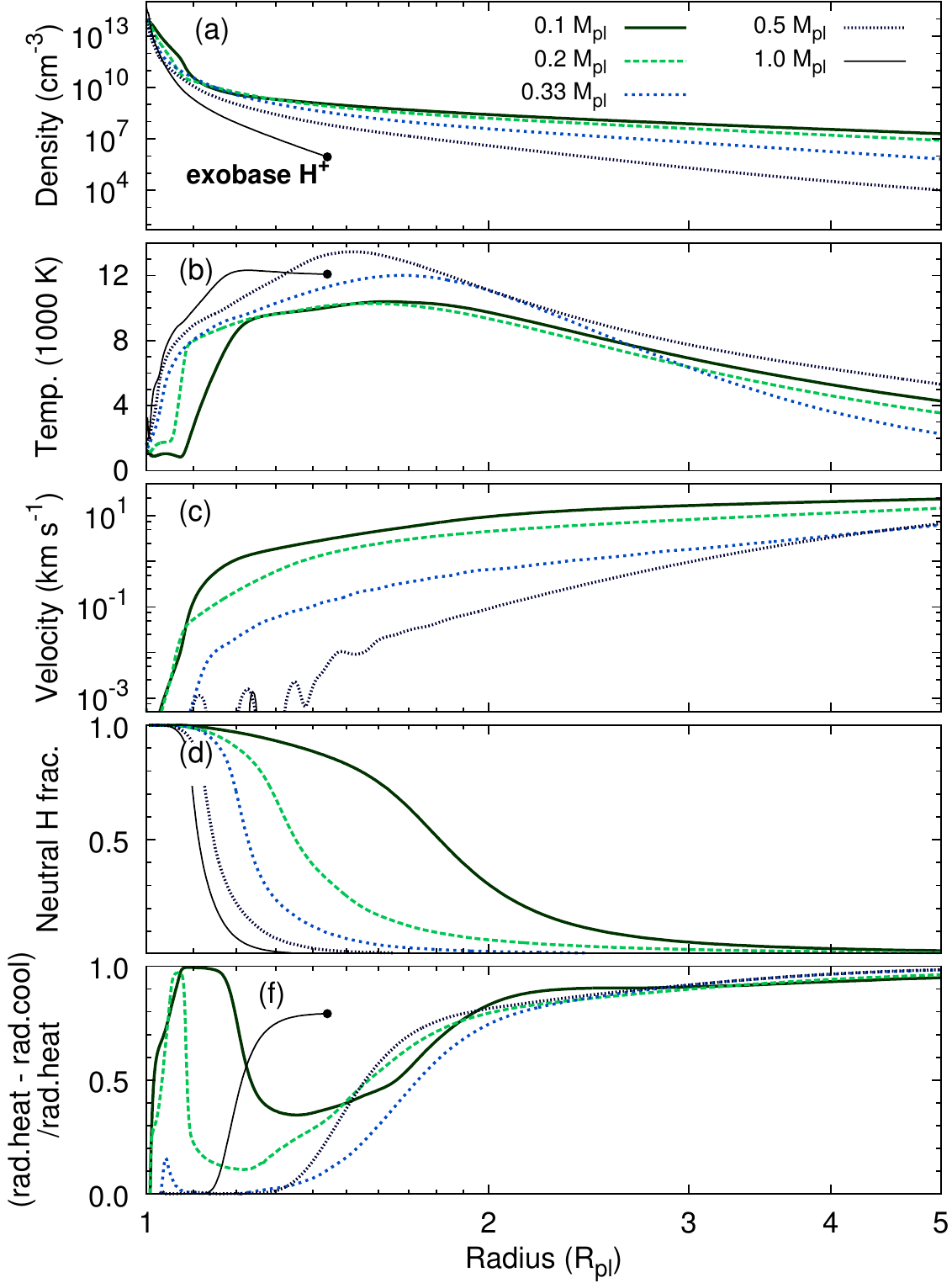}
  \caption{Atmospheres of artificial planets with different 
           planetary masses placed at the distance
           of WASP-10\,b.
           For the planet WASP-10\,b with the true mass,
           we plot the atmospheric structure up to the exobase.
           }
  \label{fig:wasp10_inc_mass}
\end{figure}

To study the dependence of the hydrodynamic winds on the planetary gravitational potential, we performed six new simulations. We considered artificial systems based on HD\,209458 and WASP-10 by adopting various masses for the planets, but keeping all other parameters fixed (see Table~\ref{tabSim}). The resulting atmospheric structures for the systems based on WASP-10 are plotted in Fig.~\ref{fig:wasp10_inc_mass}.
As we go from  low to high planetary masses, the density and wind velocity in the thermosphere decrease, while the temperature increases. The higher temperature in the thermospheres of more massive planets causes enhanced radiative cooling, which is dominated by \lya{} cooling above $\sim$1.1~$R_{\mathrm{pl}}$ and by free-free emission below. This radiative cooling reduces the heating efficiency (see Fig.~\ref{fig:wasp10_inc_mass} (f)), which limits the available energy for accelerating the planetary wind. The lower wind velocity causes less adiabatic cooling, which further increases the temperature in the thermosphere. It also advects less neutral hydrogen into the upper thermospheres and thereby shifts the transition height from H to H$^+$ closer to the planetary photosphere (see Fig.~\ref{fig:wasp10_inc_mass} (d)). Hence, massive planets have hot, highly ionized thermospheres that re-emit large parts of the radiative energy input and, therefore, produce weaker winds.

The atmosphere of WASP-10\,b with the true planetary mass of $M_{\mathrm{pl}}=3.2~M_{\mathrm{jup}}$ is collisionless above 1.44~$R_{\mathrm{pl}}$. The acceleration below the exobase is insufficient to drive a transonic wind, and our simulation is not valid above the exobase. By computing the mass-loss rate according to Jeans escape at the exobase \citep{Lecavelier2004}, we derive a value of only 0.6~g\,s$^{-1}$. We call thermospheres of planets with such negligibly low thermal mass-loss rates ``stable''.

%_______________________________________________________________________________
%_______________________________________________________________________________
\vspace{-2mm}
\section{Determining the free parameters in the energy-limited escape formula}\label{SectNewErgLim}
%_______________________________________________________________________________
%_______________________________________________________________________________

The energy-limited escape formula can only yield reasonable estimates if the heating efficiency and the atmospheric expansion radius are adapted to the properties of the planet under consideration. We now compute the expansion radii and heating efficiencies for all simulations presented by \citet{Salz2015c} plus our new simulations and derive scaling laws for the two factors. The resulting values are provided in Table~\ref{tabSim}.

%_______________________________________________________________________________
\vspace{-3mm}
\subsection{Atmospheric expansion}\label{SectR_XUV}
%_______________________________________________________________________________

We calculate the XUV absorption radii $R_{\mathrm{XUV}}$ in the simulations following \citet{Erkaev2007}:
\vspace{-1mm}
\begin{equation}\label{Eq:Rxuv}
  R_{\mathrm{XUV}} = \left(
                     \frac{\int^{R_\mathrm{Rl}}_{R_\mathrm{pl}} r^2 h(r) \,dr}
                          {F'_{\mathrm{XUV}}}\right)^{1/2}
                     \quad\mathrm{with}\quad
                     F'_{\mathrm{XUV}} = 
                     \int^{R_\mathrm{Rl}}_{R_\mathrm{pl}} h(r) \,dr .
\end{equation}

\vspace{-1mm}\noindent Here, $R_\mathrm{Rl}$ is the Roche-lobe height and $h(r)$ the local heating rate. $F'_{\mathrm{XUV}}$ holds the fraction of the radiative energy input that heats the atmosphere; this means that energy used for ionizing hydrogen,
for example, is excluded. Mathematically, the total radiative heating produced by the absorption of the complete XUV flux in a thin layer at $R_{\mathrm{XUV}}$ is equivalent to the distributed heating in the simulations. 

The resulting $R_{\mathrm{XUV}}$ increase with decreasing gravitational potential and, to a lesser extent, a higher irradiation level causes the thermospheres to expand. We use a singular value decomposition (SVD) to fit a power law to the dependency of $\beta = R_{\mathrm{XUV}}/R_{\mathrm{pl}}$ on the gravitational potential, $\Phi_{\mathrm{G}} = -G M_\mathrm{pl}/R_\mathrm{pl}$, and the irradiating XUV flux, $F_{\mathrm{XUV}}$:
\begin{align}\label{Eq:fit_Rxuv}
  \notag&\log_{10}\left( \beta\right) = \max\big(0.0,\\
  &-0.185\,\log_{10}\left( -\Phi_{\mathrm{G}}\right)
                    +0.021\,\log_{10}\left( F_{\mathrm{XUV}} \big)
                    +2.42 \right).
\end{align}
The fit is shown in Fig.~\ref{fig:rxuv}. The atmospheric expansion can be neglected for massive hot Jupiters, but in the range of super-Earth-sized planets the expansion causes mass-loss rates that are higher by a factor of four.

%_______________________________________________________________________________
\vspace{-3mm}
\subsection{Evaporation efficiency}
%_______________________________________________________________________________

We derive an evaporation efficiency in the individual simulations by comparing the energy-limited mass-loss rates with the simulated mass-loss rates 
\begin{equation}\label{Eq:eta_eff}
  \eta_{\mathrm{eva}} = \dot{M}_{\mathrm{sim}}/\dot{M}_{\mathrm{el}} .
\end{equation}
While the conventional heating efficiency $\eta$ specifies the fraction of the radiative energy input that is converted into heat, the \emph{\textup{evaporation}} efficiency only accounts for the energy converted into gravitational potential energy \citep{Lopez2012}. Therefore, the use of $\eta_{\mathrm{eva}}$ in Eq.~\ref{EqMloss_erglim} compensates for neglecting the kinetic and thermal energy gain in the energy-limited approach and reconciles the derived mass-loss rates with the fundamental energy budget consideration introduced in the beginning of Sect.~\ref{SectErgLimEsc}. On average, we find $\eta = 5/4\;\eta_{\mathrm{eva}}$ in our simulations.

The mass-loss rates are evaluated according to $\dot{M}_{\mathrm{sim}}= 1/4 \hspace{0.05em}\times\hspace{0.05em} 4\pi R^2\,\rho\mathrm{v}$; the additional factor of $1/4$ results from averaging the irradiation over the planetary surface. The energy-limited mass-loss rates are calculated from Eq.~\ref{EqMloss_erglim} with a heating efficiency of $\eta = 1.0$ \citep[system parameters are provided in Table~1 of][]{Salz2015c}. We use the $\beta$ factors from the previous section for the energy-limited escape rates, which ensures that the energy budget is the same in both mass-loss evaluations. 

\begin{figure}
    \centering
    \includegraphics[width=\hsize]{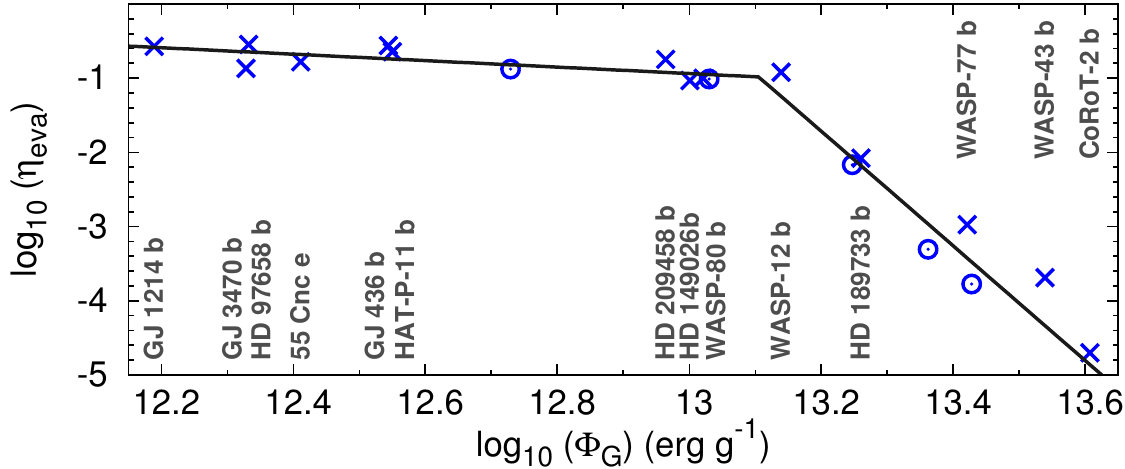}
    \caption{Evaporation efficiencies in the simulations.
             The solid line shows our fit.
             Planets are labeled and artificial planets
             are marked by circles.
           }
    \label{fig:eta_corr}
\end{figure}

Figure~\ref{fig:eta_corr} shows the evaporation efficiencies plotted versus the gravitational potential. The resulting evaporation efficiencies reach at most 0.29 (HD\,97658\,b), and the most compact planet with a wind has an efficiency of only $10^{-4.7}$ (CoRoT-2\,b). While the evaporation efficiency is almost constant for planets with $\log_\mathrm{10}\left( -\Phi_{\mathrm{G}}\right) < 13.11$~erg\,g$^{-1}$, it rapidly declines for planets with higher gravitational potentials. Fitting this dependency with a broken power law and using $v = \log_\mathrm{10}\left( -\Phi_{\mathrm{G}}\right)$, we obtain
\vspace{-1mm}
\begin{align}\label{Eqeta_correct}
  \log_\mathrm{10}\left( \eta_{\mathrm{eva}}\right) = 
  \begin{cases}
    -0.50 - 0.44 \left(v - 12.0\right)  & \text{for }  12.0 < v \le 13.11,\\
    -0.98 - 7.29 \left(v - 13.11\right)  & \text{for }  13.11 < v < 13.6 .
  \end{cases}
\end{align}

\vspace{-1mm}\noindent Planets with $\log_\mathrm{10}\left( -\Phi_{\mathrm{G}}\right) \gtrsim 13.6$~erg\,g$^{-1}$ have stable thermospheres and are excluded from the fit. For smaller planets up to HAT-P-11\,b, the evaporation efficiencies show almost no trend, and the use of a constant evaporation efficiency of $\eta_{\mathrm{eva}} = 0.23$ is a reasonable approximation.

%_______________________________________________________________________________
\vspace{-3mm}
\subsection{Approximations, comparisons, and range of validity}\label{SectValid}
%_______________________________________________________________________________

According to our definition, all parameters that affect the planetary mass-loss rates introduce uncertainties in the calculation of $\eta_{\mathrm{eva}}$. For example, up to 10\% of the XUV energy input is transmitted through the lower boundary (X-rays up to $\approx$\,20~\AA{}). In contrast, we also find small amounts of hydrogen line heating, especially \lya{} line heating in the denser atmosphere close to the lower boundary. Additionally, our boundary conditions slightly affect the mass-loss rates. The impact of metals and molecules on the mass-loss rates was studied in two test simulations of HD\,209458\,b and HD\,189733\,b by including metals with solar abundances \citep{Salz2015c}. The combined relative uncertainty introduced by these factors is about a factor of two \citep[see Table~2 in][]{Salz2015c}, while the presented evaporation efficiencies vary by several orders of magnitude.

The atmosphere of HD\,209458\,b shows $\eta_{\mathrm{eva}}=0.18,$ and corrected for the kinetic and thermal energy gain, we derive $\eta=0.21$. This result is slightly higher than the heating efficiency of $0.1\le\eta<0.15$ derived by \citet{Shematovich2014}. These authors also simulated the absorption of XUV radiation in the lower thermosphere of HD\,209458\,b including H and He but without solving the dynamics of the gas. 
Our heating efficiency is higher because it is an average over the complete thermosphere up to the Roche-lobe height, whereas \citeauthor{Shematovich2014} focused on the lower thermosphere, where the heating efficiency is lower (see Fig.~\ref{fig:wasp10_inc_mass}~(f)). They found collisional excitations and ionization through thermal and suprathermal electrons to be an important energy sink. Our photoionization solver includes these collisional processes assuming local energy deposition of the photoelectrons \citep[details in][]{Salz2015c, Ferland1998, Ferland2013}.

Compared to the calculation of \citeauthor{Shematovich2014}, our simulation is based on a different atmospheric structure of HD\,209458\,b, and in addition, we include free-free emission of thermal electrons, while they included H$_2$ dissociation and rovibrational excitations. Therefore, the two simulations are not completely comparable (see App.~\ref{Sect:heff}). Finally, in a test simulation including molecules, H$^-$ has been shown to be a significant energy sink \citep{Salz2015c}, but it is neither included in the presented simulation nor in the scheme of \citet{Shematovich2014}. Lacking a complete treatment of all potentially relevant processes, our heating efficiencies represent an improvement over previous estimates, but not the final answer. The fact that the heating efficiencies derived by \citeauthor{Shematovich2014} and us are similar supports the validity of these results, which are based on different numerical methods.
Recent studies based on Kepler data also yield an evaporation efficiency of about 10\% for smaller hot gas planets \citep{Lopez2014}. However, given the uncertainties in the mass-loss rates especially at young ages ($<100$~Ma), this similarity could be coincidental.

\begin{figure}
    \centering
    \includegraphics[width=\hsize]{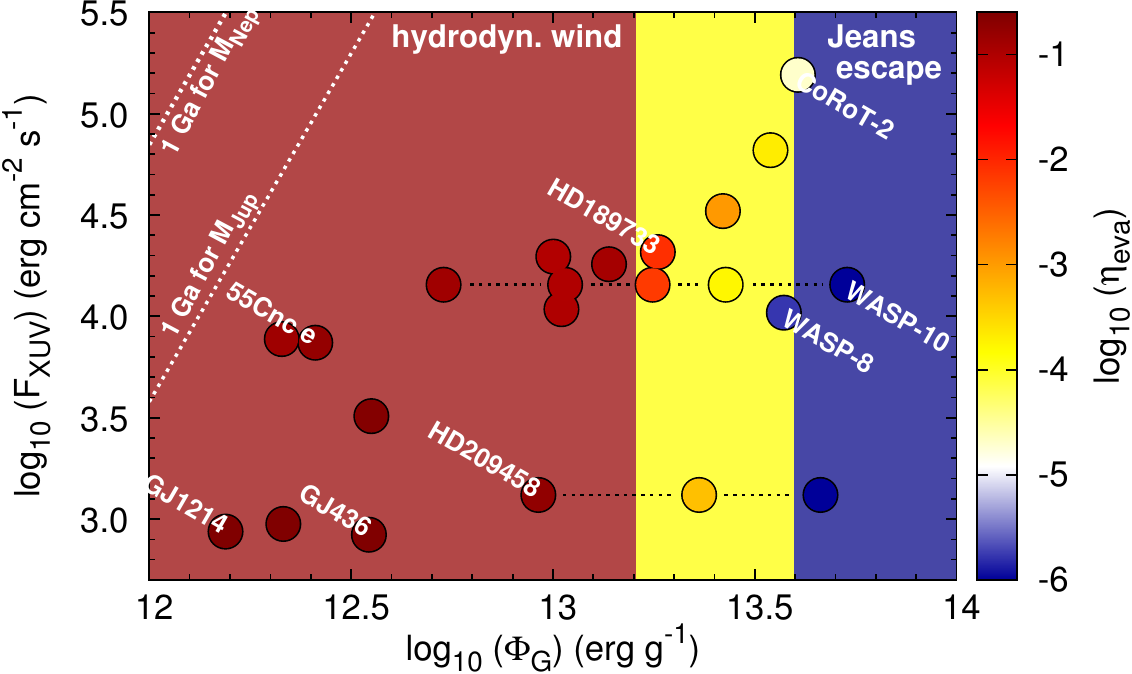}
     \caption{System parameters covered by our simulations.
              The evaporation efficiency is referenced by the 
              color scheme.
              We indicate regions of strong planetary winds,
              intermediate winds, and stable atmospheres with
              red, yellow, and blue background colors.
              The dotted lines show 1~Ga evaporation borders
              for Jupiter- and Neptune-mass planets.
              The new series of simulations are indicated by
              black dashed lines.
              }
    \label{fig:eta_irrad} 
\end{figure}

In Fig.~\ref{fig:eta_irrad} we show the system parameters covered by our simulations, which indicates the validity range of our scaling laws. A planetary wind does not exist for planets with a gravitational potential energy in excess of $\log_\mathrm{10}\left( -\Phi_{\mathrm{G}}\right) \gtrsim 13.6$~erg\,g$^{-1}$ independent of the irradiation level. In an intermediate region from $13.2 < \log_\mathrm{10}\left( -\Phi_{\mathrm{G}}\right) < 13.6$~erg\,g$^{-1}$ hydrodynamic escape can exist, but is easily suppressed by the stellar wind pressure over large portions of the planetary surface \citep{Salz2015c}. None of the simulated planets is currently unstable because of its mass-loss rate, as indicated by the evaporation borders.

%_______________________________________________________________________________
%_______________________________________________________________________________
\vspace{-1mm}
\section{Summary and conclusion}
%_______________________________________________________________________________
%_______________________________________________________________________________

Based on TPCI simulations of the escaping thermospheres of hot gas planets, we found a transition from strong winds in planets with a low gravitational potential to stable atmospheres in massive and compact planets. This transition is the result of strong radiative cooling in the hot thermospheres of massive hot Jupiters. Planets with a gravitational potential higher than $\log_\mathrm{10}\left( -\Phi_{\mathrm{G}}\right) \gtrsim 13.6$~erg\,g$^{-1}$ cannot host hydrodynamically escaping atmospheres, but planets with $\log_\mathrm{10}\left( -\Phi_{\mathrm{G}}\right) \leq 13.11$~erg\,g$^{-1}$ show mass-loss rates that are close to being energy limited. Furthermore, the mass-loss rates of smaller planets are additionally increased by significantly expanded atmospheres. We provided scaling laws for the decreasing heating efficiencies and for the expansion radii depending on the gravitational potential and the irradiation level of the planets.

In practice, the atmospheric expansion and the evaporation efficiency can be derived for any system with Eqs.~\ref{Eq:fit_Rxuv}~and~\ref{Eqeta_correct}, and the resulting values can be used in Eq.~\ref{EqMloss_erglim} to derive the planetary mass-loss rate. The revised energy-limited escape is valid for atmospheres with up to solar metal abundances if they are ``freely escaping'', namely, not under the influence of strong planetary magnetic fields or high stellar wind pressures \citep{Murray2009, Trammell2014}. Under these conditions, the formalism can be applied to planets
with masses from super-Earths to hot Jupiters. Our scaling laws incorporate the effects of strong radiative cooling in compact and hot atmospheres and the increased absorption area of the expanded atmospheres in smaller planets into the energy-limited escape concept.

\begin{acknowledgements}
We thank J. Sanz-Forcada for a useful discussion on the $\beta$ factor and the referee F. Tian for his helpful comments.
MS acknowledges support by Verbundforschung (50OR 1105) and
the German National Science Foundation (DFG) within the Research Training College 1351.
PCS acknowledges support by the DLR under 50 OR 1307 and by an ESA research fellowship.
\end{acknowledgements}

%_______________________________________________________________________________
\bibliographystyle{aa}
\setlength{\bibsep}{0.0pt}
\vspace{-5mm}
\bibliography{salz_erglim_esc}

\appendix

\section{System parameters}

The parameters are provided in Table~\ref{tabSim}.

\begin{table*}
\small
\caption{Irradiation of the planets, energy-limited mass-loss rates, and 
         results from the simulations. Ranking according to simulated mass loss.}
\label{tabSim}      
\centering          
\begin{tabular}{l @{\hspace{6pt}}l cccccc}
\hline\hline\vspace{-8pt}\\
  System &  
   & 
  $\log_\mathrm{10}\left( F_{\mathrm{XUV}}\right)$\tablefootmark{d} & 
  $\log_\mathrm{10}\left( -\Phi_{\mathrm{G}}\right)$ & 
  $R_{\mathrm{XUV}}$ & 
  $\log_\mathrm{10}\left( \dot{M}_{\mathrm{el}}\right)$ & 
  $\log_\mathrm{10}\left( \dot{M}_{\mathrm{sim}}\right)$ & 
  $\log_\mathrm{10}\left( \eta_{\mathrm{eva}}\right)$  \\ 
  \vspace{-9pt}\\
   & 
   &  
   &
  (erg\,g$^{-1}$)  &
  ($R_{\mathrm{pl}}$) &
  (g\,s$^{-1}$)  &
  (g\,s$^{-1}$) &
   \\
  \vspace{-8pt}\\ \hline\vspace{-8pt}\\ 
  WASP-12\,b           &&             $<$\,4.26 & 13.14 & \hphantom{$<$}\,1.20  &  12.52  & 11.60                              & \hphantom{$<$}\,$-0.92$  \\ 
  GJ\,3470\,b          &&  \hphantom{$<$}\,3.89 & 12.33 & \hphantom{$<$}\,1.77  &  11.53  & 10.66                              & \hphantom{$<$}\,$-0.87$  \\ 
  WASP-80\,b           &&  \hphantom{$<$}\,4.03 & 13.02 & \hphantom{$<$}\,1.24  &  11.55  & 10.55                              & \hphantom{$<$}\,$-1.00$  \\ 
  HD\,149026\,b        &&  \hphantom{$<$}\,4.29 & 13.00 & \hphantom{$<$}\,1.26  &  11.46  & 10.43                              & \hphantom{$<$}\,$-1.03$  \\ 
  HAT-P-11\,b          &&  \hphantom{$<$}\,3.51 & 12.55 & \hphantom{$<$}\,1.61  &  10.93  & 10.29                              & \hphantom{$<$}\,$-0.64$  \\ 
  HD\,209458\,b        &&             $<$\,3.06 & 12.96 & \hphantom{$<$}\,1.25  &  11.02  & 10.27                              & \hphantom{$<$}\,$-0.75$  \\ 
  55\,Cnc\,e           &&  \hphantom{$<$}\,3.87 & 12.41 & \hphantom{$<$}\,1.59  &  10.92  & 10.14                              & \hphantom{$<$}\,$-0.78$  \\ 
  GJ\,1214\,b          &&  \hphantom{$<$}\,2.93 & 12.19 & \hphantom{$<$}\,1.51  &  10.25  & \hphantom{1}9.68                   & \hphantom{$<$}\,$-0.57$  \\ 
  GJ\,436\,b           &&  \hphantom{$<$}\,2.80 & 12.54 & \hphantom{$<$}\,1.48  &  10.21  & \hphantom{1}9.65                   & \hphantom{$<$}\,$-0.56$  \\ 
  HD\,189733\,b        &&  \hphantom{$<$}\,4.32 & 13.26 & \hphantom{$<$}\,1.15  &  11.69  & \hphantom{1}9.61                   & \hphantom{$<$}\,$-2.08$  \\ 
  HD\,97658\,b         &&  \hphantom{$<$}\,2.98 & 12.33 & \hphantom{$<$}\,1.75  &  10.01  & \hphantom{1}9.47                   & \hphantom{$<$}\,$-0.54$  \\ 
  WASP-77\,b           &&  \hphantom{$<$}\,4.51 & 13.42 & \hphantom{$<$}\,1.10  &  11.76  & \hphantom{1}8.79                   & \hphantom{$<$}\,$-2.97$ \\ 
  WASP-43\,b           &&  \hphantom{$<$}\,4.82 & 13.54 & \hphantom{$<$}\,1.08  &  11.73  & \hphantom{1}8.04                   & \hphantom{$<$}\,$-3.69$  \\ 
  Corot-2\,b           &&  \hphantom{$<$}\,5.19 & 13.61 & \hphantom{$<$}\,1.06  &  12.33  & \hphantom{1}7.63                   & \hphantom{$<$}\,$-4.70$  \\ 
  WASP-8\,b            &&  \hphantom{$<$}\,3.66 & 13.57 & \hphantom{$<$}\,1.07  &  10.77  & \hphantom{1}{\it 5.0}\hphantom{1}  & \hphantom{$<$}\,{\it $-$5.77}  \\ 
  WASP-10\,b           &&  \hphantom{$<$}\,4.08 & 13.73 & \hphantom{$<$}\,1.07  &  10.86  & \hphantom{1}{\it 2.7}\hphantom{1}  & \hphantom{$<$}\,{\it $-$8.16}  \\                                                                                                                                              
  HAT-P-2\,b           &&  \hphantom{$<$}\,4.12 & 14.14 &            $<$\,1.05  &  10.83  & {\it \!\!$<$\,5.9}\hphantom{1}     & {\it $< -$5.4}\hphantom{1} \\                                                                                                                                            
  HAT-P-20\,b          &&  \hphantom{$<$}\,4.08 & 14.18 &            $<$\,1.05  &  10.33  & {\it \!\!$<$\,4.5}\hphantom{1}     & {\it $< -$6.3}\hphantom{1}  \\ 
  \vspace{-9pt}\\
  \multicolumn{8}{c}{\it artificial planets}\\
  \vspace{-9pt}\\
  HD\,209458\,b (2.5)  &&  \hphantom{$<$}\,3.06 & 13.36 & \hphantom{$<$}\,1.10  &  10.45  & \hphantom{1}7.15                   & \hphantom{$<$}\,$-3.31$  \\ 
  HD\,209458\,b (5.0)  &&  \hphantom{$<$}\,3.06 & 13.66 & \hphantom{$<$}\,1.04  &  10.07  & {\it \!\!$<$\,2.0}\hphantom{1}  & {\it $< -$8.1}\hphantom{1}   \\ 
  WASP-10\,b    (0.1)  &&  \hphantom{$<$}\,4.08 & 12.73 & \hphantom{$<$}\,1.42  &  12.23  & 11.35                              & \hphantom{$<$}\,$-0.88$  \\  
  WASP-10\,b    (0.2)  &&  \hphantom{$<$}\,4.08 & 13.03 & \hphantom{$<$}\,1.25  &  11.77  & 10.76                              & \hphantom{$<$}\,$-1.01$  \\  
  WASP-10\,b    (0.33) &&  \hphantom{$<$}\,4.08 & 13.25 & \hphantom{$<$}\,1.16  &  11.46  & \hphantom{1}9.27                   & \hphantom{$<$}\,$-2.17$  \\  
  WASP-10\,b    (0.5)  &&  \hphantom{$<$}\,4.08 & 13.43 & \hphantom{$<$}\,1.10  &  11.21  & \hphantom{1}7.44                   & \hphantom{$<$}\,$-3.77$  \\  
  \vspace{-9pt}\\ \hline
\end{tabular}                                                                                                                                                                    
\tablefoot{Explanation of the columns: 
           name of the system
           (mass scaling of artificial planets),
           XUV flux at planetary distance \tablefoottext{d}{($<$\,912~\AA{}, erg\,cm$^{-2}$\,s$^{-1}$, see \citet{Salz2015c})},
           gravitational potential of the planet,
           bulk XUV absorption radius in the simulations,
           energy-limited mass-loss rate \citep{Sanz2011},
           mass-loss rate in the simulations 
           ($1/4 \hspace{0.05em}\times\hspace{0.05em} 4\pi R^2\,\rho\mathrm{v}$),
           evaporation efficiency (see text).
           Mass-loss rates and evaporation efficiencies of 
           hydrodynamically stable atmospheres are printed in italics.
           }
\end{table*}                                                                                                                                                                     

\vspace{-2mm}
\section{Atmospheric expansion}

\begin{figure}[h]
    \centering
    \includegraphics[width=\hsize, trim=0cm 0cm 0cm 0cm, clip=true]{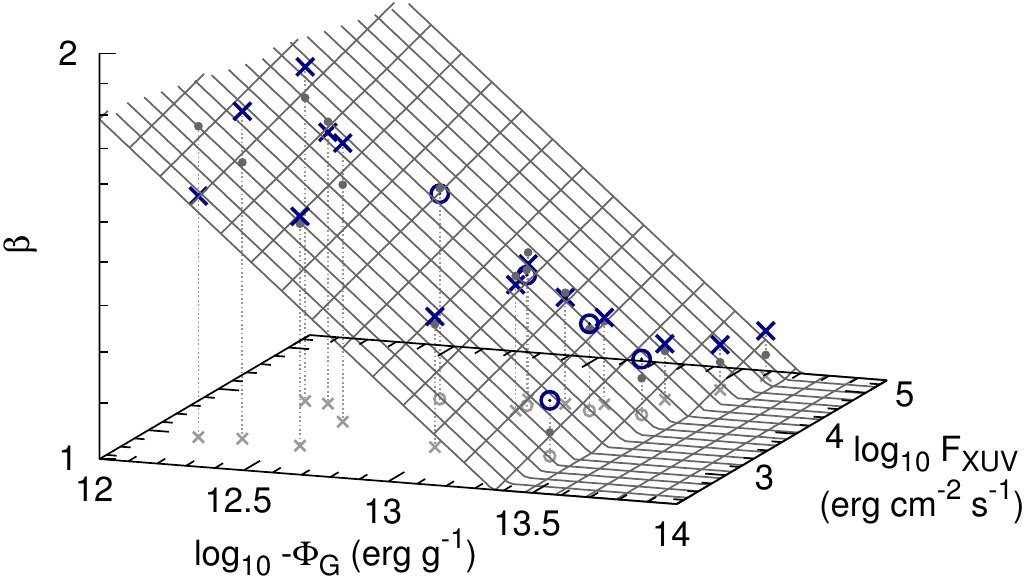}
    \caption{Mean XUV absorption radii in the simulations.
             Our fit is given by the plane.
             Exoplanetary systems are indicated by crosses and
             artificial planets by circles. }
    \label{fig:rxuv}
\end{figure}

\vspace{-5mm}
\section{Heating efficiency}\label{Sect:heff}

\begin{figure}[h]
    \centering
    \includegraphics[width=0.9\hsize, trim=0cm 0cm 0cm 0cm, clip=true]{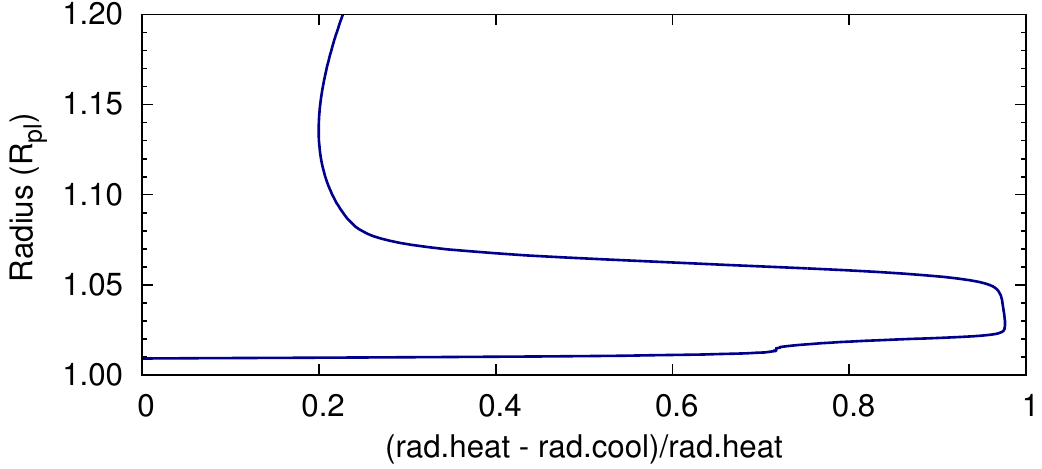}
    \caption{Heating fraction plotted versus the planetary radius in
             the atmosphere of HD\,209458\,b.
             }
    \label{fig:heff}
\end{figure}

In Fig.~\ref{fig:heff} we show the structure of the heating fraction in the atmosphere of HD\,209458\,b, which can be compared to Fig.~4 of \citet{Shematovich2014}. We note that we plot the heating fraction, which differs from the heating efficiency mainly by the fraction of the radiative energy input used for ionization processes. Nevertheless, the general height structure can be compared. 
The heating fraction is small below $1.02~R_{\mathrm{pl}}$, which is caused by free-free emission of thermal electrons, and the dip above $1.05~R_{\mathrm{pl}}$ is caused by hydrogen \lya{} cooling. \citet{Shematovich2014} also found a maximum in the heating efficiency around  $1.08~R_{\mathrm{pl}}$. The exact position of this maximum depends on the density and temperature structure of the atmosphere, which is different in the two compared simulations. The authors did not explain the main processes for atmosperic cooling above or below the peak, so it is not clear whether the decrease in the heating efficiency above $1.08~R_{\mathrm{pl}}$ corresponds to the \lya{} cooling found in our simulations.
Certainly, their finding of a lower heating efficiency below $1.08~R_{\mathrm{pl}}$ is not based on free-free emission of thermal electrons as in our case because their scheme does not include this process. We suspect molecular processes to cause this cooling in their simulations. A more detailed comparison can only be achieved by using the same atmospheric structure.

\end{document}